\documentclass[aps,reprint,superscriptaddress]{revtex4-1}

\usepackage{graphicx,color}
\usepackage{graphics}
\usepackage{lipsum}
\usepackage{float}
\floatstyle{boxed}
\usepackage{blkarray}
\usepackage{flushend}

\begin{document}


\title{Assessment of a silicon quantum dot spin qubit environment via noise spectroscopy}


\author{K. W. Chan}
\email{ckwai85@gmail.com}
\author{W. Huang}
\author{C. H. Yang}
\author{J. C. C. Hwang}
\author{B. Hensen}
\author{T. Tanttu}
\author{F. E. Hudson}
\affiliation{Centre for Quantum Computation and Communication Technology, School of Electrical Engineering and Telecommunications, The University of New South Wales, Sydney NSW 2052, Australia}
\author{K. M. Itoh}
\affiliation{School of Fundamental Science and Technology, Keio University, 3-14-1 Hiyoshi, Kohoku-ku, Yokohama 223-8522 Japan}
\author{A. Laucht}
\author{A. Morello}
\author{A. S. Dzurak}
\email{a.dzurak@unsw.edu.au}
\affiliation{Centre for Quantum Computation and Communication Technology, School of Electrical Engineering and Telecommunications, The University of New South Wales, Sydney NSW 2052, Australia}


\date{\today}

\begin{abstract}
	
Preserving coherence long enough to perform meaningful calculations is one of the major challenges on the pathway to large scale quantum computer implementations. Noise coupled in from the environment is the main contributing factor to decoherence but can be mitigated via engineering design and control solutions. However, this is only possible after acquiring a thorough understanding of the dominant noise sources and their spectrum. In this paper, we employ a silicon quantum dot spin qubit as a metrological device to study the noise environment experienced by the qubit. We compare the sensitivity of this qubit to electrical noise with that of an implanted silicon donor qubit in the same environment and measurement set-up.  Our results show that, as expected, a quantum dot spin qubit is more sensitive to electrical noise than a donor spin qubit due to the larger Stark shift, and the noise spectroscopy data shows pronounced charge noise contributions at intermediate frequencies (2--20 kHz). 


\end{abstract}


\maketitle


Spin-based quantum dot qubits~\cite{LossDiVincenzo1998_PhysRevA.57.120} in semiconductors show promise for scalable quantum information processing due to their compatibility with well-established semiconductor manufacturing technologies. Extremely long spin coherence times have been demonstrated in spin qubits fabricated on isotopically purified silicon~\cite{muhonen2014storing,Veldhorst_Natnano2014,eng2015isotopically,tyryshkin2012electron}, with control and readout fidelities exceeding fault-tolerance thresholds~\cite{Veldhorst_Natnano2014, yoneda2018quantum}. Two-qubit logic gates~\cite{Veldhorst_Nature2015, watson2018programmable, zajac2018resonantly, huang2018fidelity} based on silicon quantum dots have also been demonstrated as a consequence of these advancements. Scaling up to larger multi-qubit systems, however, requires a more stringent engineering of the qubits' electromagnetic environment such that the collective fault-tolerant threshold is maintained for the implementation of surface code error-correction protocols~\cite{jones2016logical}. This demands a detailed understanding of the possible sources of noise that cause decoherence at the very least.

Noise spectroscopy is a valuable and necessary tool in building understanding of the noise sources present. As part of the effort towards scaling up qubit systems this routine has been undertaken for superconducting~\cite{bylander2011noise}, ion trap~\cite{almog2016dynamic} and diamond NV center~\cite{PhysRevLett.114.017601} qubits. Noise spectroscopy for spin-based quantum computing in silicon has been done for an implanted phosphorus donor qubit in silicon (Si:P)~\cite{muhonen2014storing} and a SiGe quantum dot~\cite{yoneda2018quantum} spin qubit. Here, we employ a silicon metal--oxide--semiconductor (SiMOS) quantum dot spin qubit as a probe to enable noise spectroscopy via CPMG dynamical decoupling pulse sequences~\cite{alvarez2011measuring,yuge2011measurement}. We start by comprehensive characterization of the qubit which includes coherence time measurements and randomized benchmarking of the single qubit Clifford gate control fidelities.


\begin{figure}
	\includegraphics[width=\columnwidth]{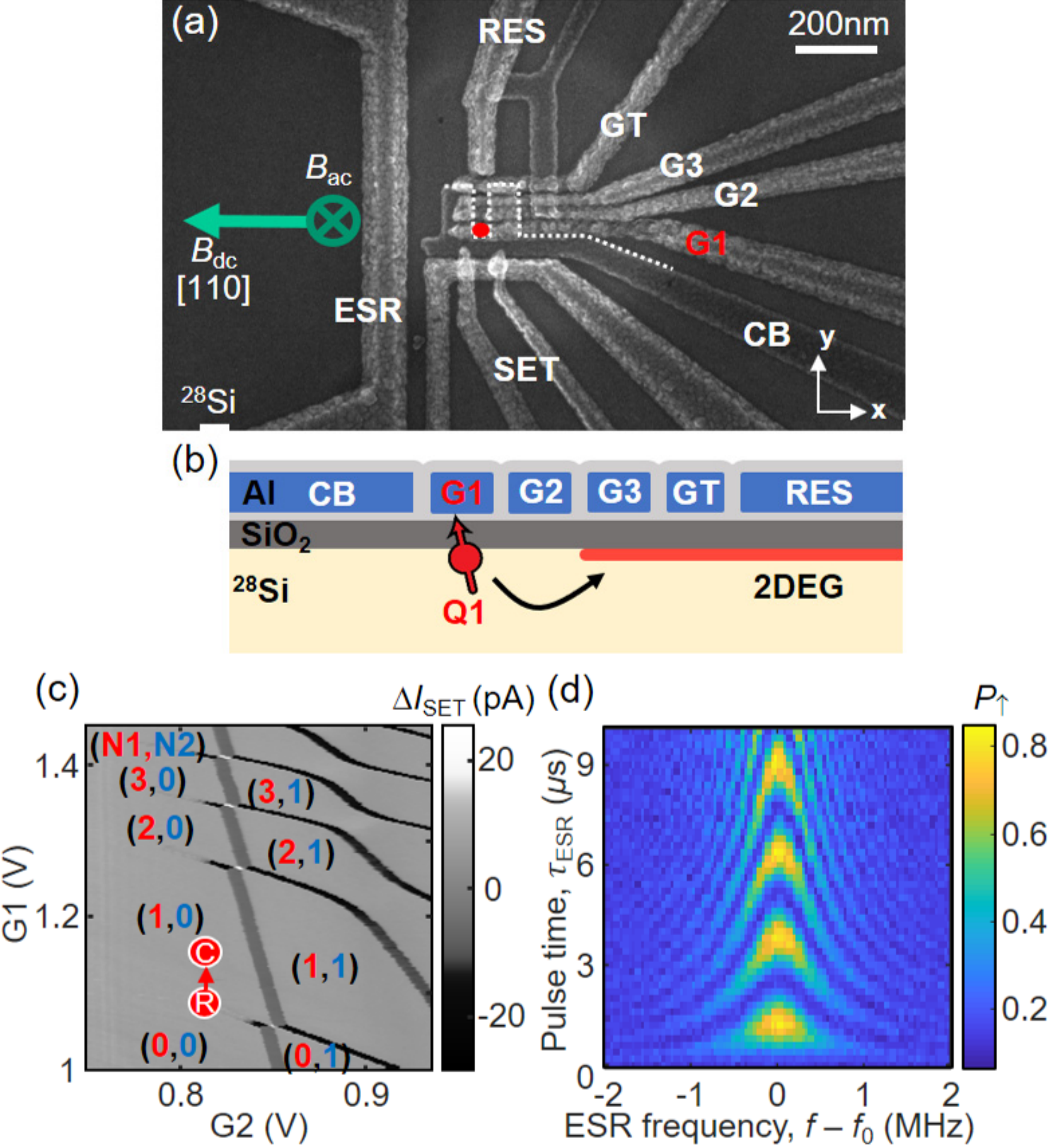}
	\caption{(a) Scanning electron micrograph of an identical SiMOS qubit device to the one under study here. CB, marked with the white dotted line is the quantum dots confinement gate. Each quantum dot is confined in a 40 nm $\times$ 40 nm area underneath of gates G1--G3. (b) Schematic cross-section of panel (a) along the y-axis of the qubit marked with a red dot (not to scale). The red region underneath gate RES illustrates a 2DEG formed with positive bias voltage, and extends to a nearby phosphorus doped ohmic region. In this paper, we report on the data obtained from qubit Q1, formed underneath of gate G1, as depicted by the red dot. (c) Charge stability diagram of a double quantum dot system confined under gates G1 and G2. The double dot electron occupancies in each of the Coulomb blockaded regions are labeled as [N1 (red), N2 (blue)] with N1 (N2) representing the number of electrons in dot G1 (G2). Here, we operate the qubit in the (0,0)--(1,0) electron configuration. The control (C) and readout (R) positions are labeled in red. (d) Rabi-chevron map showing qubit spin--up probability as a function of electron spin resonance (ESR) detuning frequency, $f-f_0$, and ESR pulse time, $\tau_{\rm ESR}$. Here, the ESR frequency is $f_{0}$ = 38.7765~GHz, $B_{\rm dc}$ = 1.4~T, and applied source microwave power, $P_{\rm ESR}$ = 5~dBm. From these results, we extract the electron Land$\acute{e}$ $g$-factor to be 1.9789.}
	\label{FIG1}
\end{figure}

Figure~\ref{FIG1}(a) shows the scanning electron micrograph (SEM) image of an identical device, fabricated on an isotopically enriched 900~nm $^{28}\textrm{Si}$ epilayer~\cite{Kohei28SiMRS2014} with an 800~ppm residual concentration of $^{29}\textrm{Si}$. This device is fabricated based on our previously reported aluminium gate stacked architecture~\cite{AngusNano07,lim2009observation}, with the distinction of employing bilayer PMMA/copolymer resist to ease the metal liftoff process. The single-electron transistor (SET) is a charge sensor~\cite{morello2010single} used to read out the charge occupancy and electron spin state of the confined quantum dots under gates G1--G3. Gate GT acts as a tunnel barrier for the loading of electrons into the quantum dots from the reservoir gate (RES). Gates G1--G3 are used to tune the electron occupancies.  A d.c.~magnetic field, $B_{\rm dc}$ of 1.4~T is applied to Zeeman-split the electron spin states to form the qubit eigenstates. The electron spin state is manipulated by utilizing the ESR microwave line to produce a perpendicular a.c.~magnetic field, $B_{\rm ac}$ at microwave frequency, $f_0$. The directions of both magnetic fields $B_{\rm dc}$ and $B_{\rm ac}$ are annotated in Fig.~\ref{FIG1}(a). Figure~\ref{FIG1}(b) shows the schematic cross-section of the device along the y-axis of the qubit, marked with a red dot in Fig.~\ref{FIG1}(a).


Figure~\ref{FIG1}(c) depicts the stability diagram showing the charge transitions on a double-dot system that is electrostatically-confined under gates G1 and G2. The electron occupancies are labeled in each Coulomb blockaded region as (N1,N2), with N1 (N2) representing the number of electrons under gates G1 (G2). In this experiment, we operate the dot under gate G1 in the (0,0)--(1,0) electron occupancy. The control (C) and readout (R) positions are labeled in red. Detailed reports on the electron spin resonance measurement technique and setup have been published in Ref.~\cite{Veldhorst_Natnano2014}. The measured Rabi-chevron pattern is depicted in Fig.~\ref{FIG1}(d). The high quality chevron shows excellent control of the electron spin, with an extracted $\pi$-pulse time of 1.28~$\mu$s. Using the single-shot spin to charge conversion technique~\cite{morello2010single,elzerman2004single}, all experimental data shown are obtained with the electron--reservoir tunnel rate tuned to $\approx 100~\mu$s with at least 100 single-shot measurements for each data point. For this qubit, we have measured a spin relaxation time $T_1~\approx1$~s and Ramsey~\cite{ramsey1990experiments} dephasing time $T_2^*$ = 33 $\pm$ 8~$\mu$s (data not shown). In addition to that, we measured the routinely-reported coherence times $T_2^{\rm H}$ = 401 $\pm$ 42~$\mu$s, $T_2^{\rm CP}$ = 1.5 $\pm$ 0.2~ms (N = 7 pulses)  and $T_2^{\rm CPMG}$ = 6.7 $\pm$ 2.9~ms (N = 122 pulses) using Hahn echo~\cite{hahn1950spin}, Carr-Purcell~\cite{carr1954effects}, and Carr-Purcell-Meiboom-Gill~\cite{meiboom1958modified} pulse sequences, respectively. 


\begin{figure}
	\includegraphics[width=0.8\columnwidth]{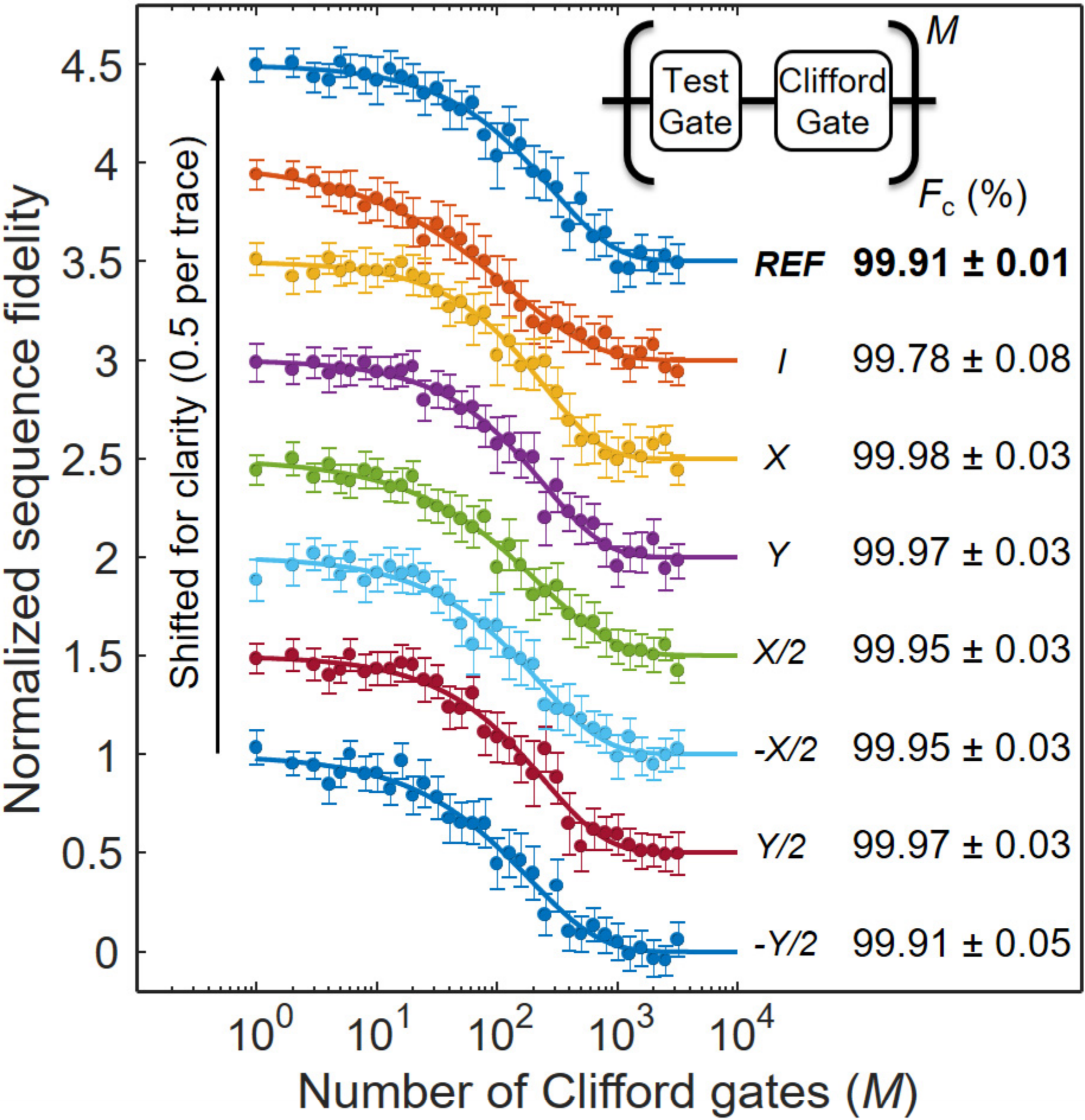}
	\caption{Randomized benchmarking of Clifford gates to determine the control fidelity of our qubit. The performance of each Clifford gate is tested by interleaving them with random Clifford gates. The Clifford gate control fidelity of this device is 99.83\%. The data are vertically shifted by 0.5 per trace for clarity. All error bars represent 95\% confidence intervals taken from the exponential fits used to extract the control fidelity.}	
	\label{RBM}
\end{figure}

\begin{figure*}
	\includegraphics[width=1\linewidth]{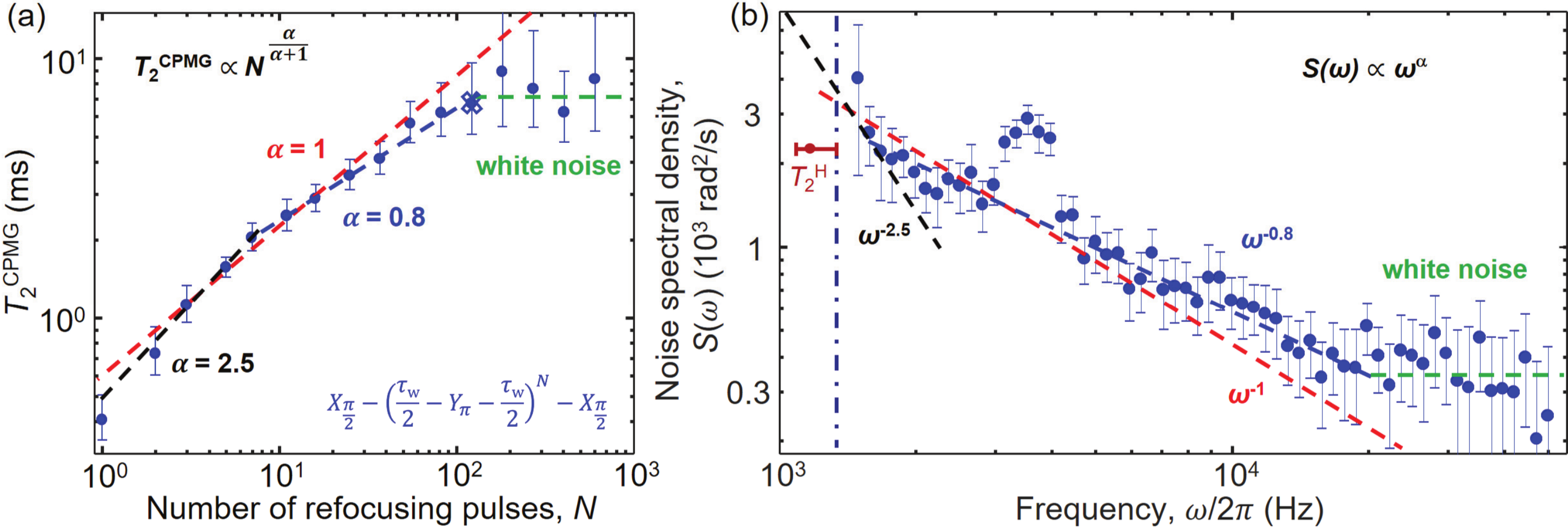}
	\caption{(a) Qubit CPMG coherence time as a function of the number of refocusing pulses, $N$. The maximum $T_2^{\rm CPMG}$ is 6.7~ms as shown in the data point marked with a cross in the plot. (b) Noise spectroscopy of a SiMOS quantum dot spin qubit. The noise power spectral density, $S(\omega)$ is calculated from the $T_2^{\rm S}$ data fitted to an exponential decay of the form, $P(t) = P_{0}exp((-t/T_2^{\rm S})^n)+P_\infty$ for different wait times, $\tau_{\rm w}$, in between the $Y_{\pi}$-pulses. 
		We observed a colored noise spectrum, with an exponent of $\alpha$ = -2.5 for $f <$ 2~kHz. In the intermediate frequencies ($f$ = 2--20~kHz), our noise is dominated by an exponent of  $\alpha$ = -0.8 to -1, very close to $1/f$ which we attribute to charge noise. We also observed a pronounced peak in the spectrum at $f \approx $ 3.6~kHz, which is caused by measurement electronics. At high frequencies ($f >$ 20~kHz), our white noise floor is $350~rad^2/$s. All error bars represent 95\% confidence intervals from the exponential fits used to extract the decay times.}
	\label{noisespec}
\end{figure*}

Next, we perform randomized benchmarking (RBM)~\cite{knill2008randomized} of Clifford gates to determine our control fidelity using standard microwave square pulses as part of the characterization. Figure~\ref{RBM} displays the extracted control fidelity (normalized) as a function of the number of Clifford gate operations, $M$. The performance of each Clifford gate is tested by interleaving them with random Clifford gates. The sequence fidelity decays over more than several 100 pulses, where $M$ is the number of Clifford gates applied. A $\pi$-pulse time of  1.75 $\mu$s and a waiting time of 100~ns between consecutive gates are used in each measurement trace. The data are vertically shifted by 0.5 per step for clarity and the visibility of all data is 0.55, limited by readout and initialization errors. The Clifford gate fidelity~\cite{magesan2012efficient} is 99.83\% which gives a primitive gate fidelity, $F_{\rm REF}$ = 99.91\%, based on the gate length 1/1.875 of the average Clifford gate length. In addition to exceeding the threshold required for quantum error correction using surface codes~\cite{hill2015surface}, this is also a factor of 4 improvement in error rate in comparison with our previous best fidelity record reported in Ref.~\cite{Veldhorst_Natnano2014}. We attribute this improvement to the utilization of the IQ modulation of a vector microwave signal generator which has a higher phase control bandwidth as opposed to the analogue phase modulation mode used in Ref.~\cite{Veldhorst_Natnano2014}. In addition to that, we have implemented ESR frequency feedback in our measurement code to keep track and correct the drift in $f_0$~\cite{blume2017demonstration}, possibly due to drift in $B_{\rm dc}$ and random charge or d.c.~voltage supply fluctuations. By achieving a high control fidelity, it is convincing that our coherence times and noise spectroscopy measurements are not limited by the ESR control pulses.


Figure~\ref{noisespec}(a) is a plot of the CPMG coherence times versus the number of refocusing pulses, $N$. The corresponding pulse sequences are shown in the bottom right text with $Y_{\pi}$ denoting a $\pi$- rotation on the y-axis of the Bloch sphere and $\tau_{\rm w}$ the wait time between the $\pi$-pulses. The coherence time can be extended by increasing $N$ until it saturates at $N$ = 122. Figure~\ref{noisespec}(b) exhibits the noise spectroscopy of our silicon quantum dot qubit. We employed our qubit as a noise probe to measure the noise power spectral densities, $S(\omega)$ using CPMG pulse sequences~\cite{alvarez2011measuring} as demonstrated for the phosphorus donor qubit system~\cite{muhonen2014storing} earlier. CPMG is a spin refocusing technique used to remove dephasing effects of low frequency transverse magnetic noise. Thus, in noise spectroscopy measurement, the CPMG sequences act as a bandpass filter, selectively choosing the portion of the noise spectrum which couples to the qubit. $S(\omega)$ is calculated from the $T_2^{\rm S}$ data fitted to an exponential decay of the form, $P(t) = P_{0}exp((-t/T_2^{\rm S})^n)+P_\infty$ for different $\tau_{\rm w}$, in between the $Y_{\pi}$-pulses. $T_2^{\rm S}$ is the electron coherence time measured while keeping $\tau_{\rm w}$ constant and progressively increasing the number of pulses in a CPMG sequence until the spin up probabilities decay completely. For each $\tau_{\rm w}$, we compute $S(\omega) = \pi^2/4T_2^{\rm S}(\omega)$ and the wait time is translated into frequency using $f = 1/2\tau_{\rm w}$ ~\cite{muhonen2014storing, yuge2011measurement}. The noise spectroscopy is limited to 1.3--50~kHz because at low frequency, $\tau_{\rm w}$ between the $Y_{\pi}$-pulses is approaching $T_2^{\rm H}$ and at high frequency, we are bound by the shortest $\tau_{\rm w}$ that is experimentally available.


In Fig.~\ref{noisespec}(b), we observed a colored noise spectrum with an exponent of $\alpha$, in close reminiscence to $\alpha$ = -2.5 for $f < 2$~kHz. The black dashed line is a plot of the function $C_{1}/\omega^{2.5}$, with $C_{1} = 3\times10^{13}$. In the Si:P donor qubit~\cite{muhonen2014storing}, this noise was attributed to the drift and fluctuations of $B_{\rm dc}$ in the superconducting magnet coils. Since both experiments are conducted in the same cryomagnetic system, the fact that their values are different ($C_{1} = 6\times10^{11}$ in Ref.~\cite{muhonen2014storing}), hints that the fluctuations of $B_{\rm dc}$ in the superconducting magnet coils may not be the cause of  $1/f^{2.5}$ noise seen in the quantum dot and Si:P donor qubit systems at low frequency. At frequencies $f$ = 2--20~kHz, our qubit noise follows the exponent of $\alpha$ = -0.8-- -1, resembling the nature of $1/f$ charge noise~\cite{bermeister2014charge,huang2014electron,paladino20141,zimmerman2014charge, yoneda2018quantum}. The red (blue) dashed line is a plot of the function $C_{2}/\omega$ ($C_{3}/\omega^{0.8}$), with $C_{2} = 3\times10^{7}$ ($C_{3} = 4\times10^{6}$). These dashed lines are guides to the eye. At high frequencies ($f > 20$~kHz), the white noise floor is $350~rad^2/$s, marked with a green dashed line. According to the power-law of the noise, the coherence time $T_2^{\rm CPMG}$ in Fig.~\ref{noisespec}(a) is expected to scale according to the noise color as $T_2 \propto N^{\alpha/(\alpha+1)}$~\cite{medford2012scaling}. The dashed lines in Fig.~\ref{noisespec}(a) are plotted with corresponding $\alpha$ values to reflect the power-law dependence of the noise spectrum. The data shows excellent agreement with the measured dependencies. 

Interestingly, in Fig.~\ref{noisespec}(b), we also observed a pronounced peak in the spectrum at $f \approx$ 3.6~kHz, which is a feature not observed in the Si:P donor experiments~\cite{muhonen2014storing}. After thorough investigation, we found that this peak is most likely caused by the d.c.~voltage sources SIM928 used to bias the qubit device. In Appendix A, we measured the noise spectrum of the SIM928 and observe a prominent peak at the exact same frequency, $f \approx$ 3.6~kHz and attribute this to be the cause of the peak observed in the noise spectroscopy in Fig.~\ref{noisespec}(b).

\begin{figure}
	\includegraphics[width=0.8\columnwidth]{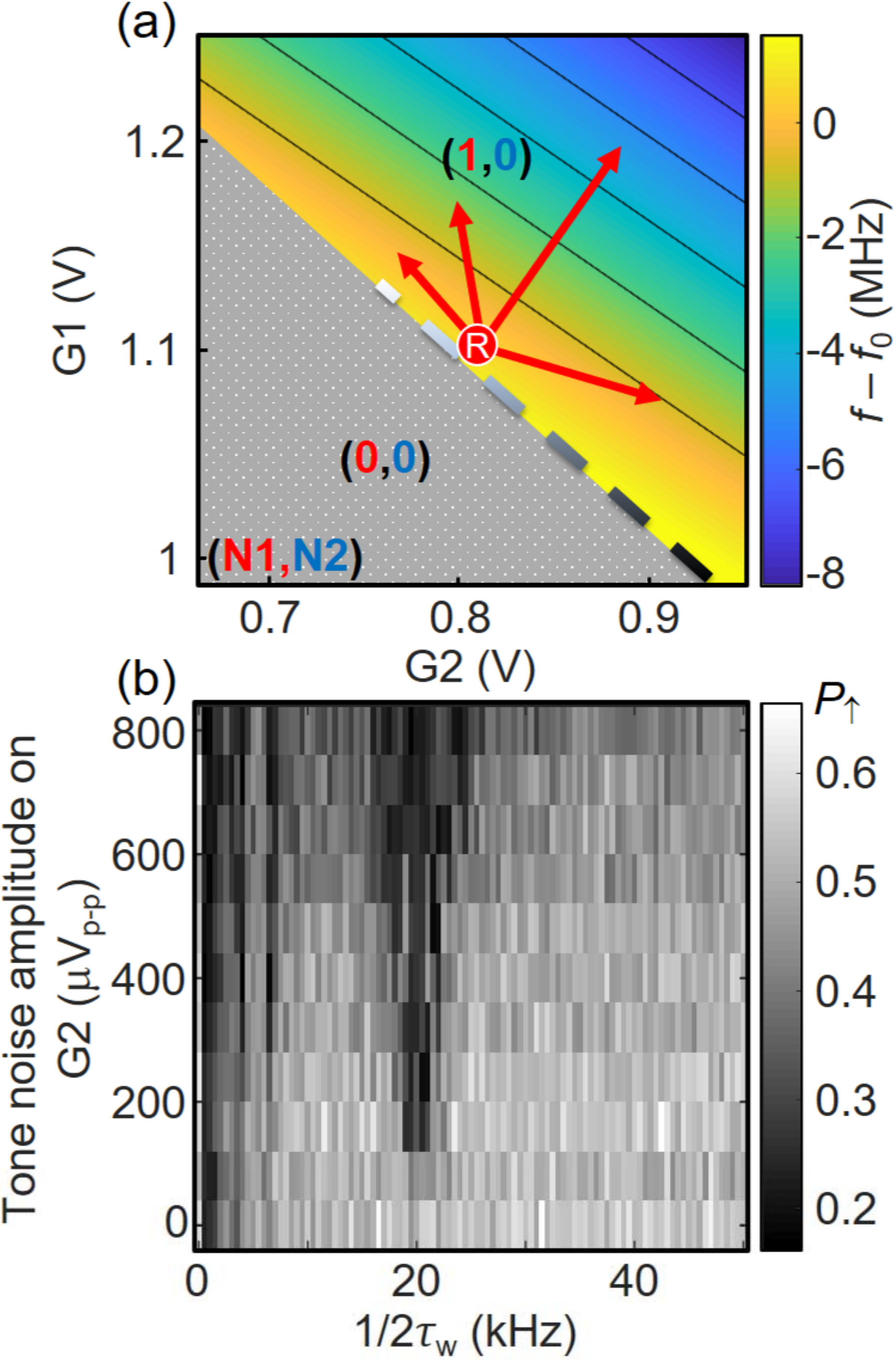}
	\caption{(a) Stark shift experienced by qubit G1 as a function of G1 and G2 gate voltages. The electron spin resonance frequencies are measured at different gate space, with 8~mV step size and extrapolated linearly as shown in the 2D map. Here, $f_{0}$ = 38.7765~GHz. From the results, we have fitted qubit G1 Stark shift to be $dg/dV_{\rm G1}$ = -36.21~MHz/V and $dg/dV_{\rm G2}$ = -22.88~MHz/V. (b) CPMG noise spectroscopy measurement while applying a $20$~kHz sinusoidal tone as a function of its amplitude on gate G2 in the y-axis. The x-axis has been translated into frequency from the CPMG wait time and all data is taken with a fixed total precession time. The results elucidate significantly lower spin up probability at the tone frequency starting from $160~\mu V_{\rm pp}$. This is a verification of our noise spectroscopy measurement technique and setup.}	
	\label{Starkshift}
\end{figure}


Figure~\ref{Starkshift}(a) shows the measurement of the Stark shift of the $g$-factor experienced by qubit G1 as a function of G1 and G2 gate voltages. From the 2D map, we extract the voltage-induced Stark shift from G1 and G2 to be $dg/dV_{\rm G1}$ = -36.21~MHz/V and $dg/dV_{\rm G2}$ = -22.88~MHz/V, respectively. Our Stark shifts are comparable to other reported silicon quantum dot qubits~\cite{Veldhorst_Natnano2014,hwang2017impact} and are much larger than the -2.27~MHz/V reported for the Si:P donor qubit~\cite{laucht2015electrically}. The enhanced Stark shift renders the quantum dot qubit more sensitive to electrical noise than Si:P donor qubit. This is obvious from $1/f$ dependence in the noise spectrum at intermediate frequencies and the higher white noise floor ($350~rad^2/$s vs.~$10~rad^2/$s for Si:P donor qubit~\cite{muhonen2014storing}). 



By applying a sinusoidal tone on gate G2 to deterministically Stark shift the qubit's frequency, we can verify our noise spectroscopy measurement technique and setup. We set the tone frequency to 20~kHz as it corresponds to the onset where $S(\omega)$ is saturated by the white noise. Figure~\ref{Starkshift}(b) is the measured qubit spin up probability after CPMG pulse sequences with different $\tau_{\rm w}$, converted into units of frequency on the x-axis, and repeated with different tone amplitudes. The results elucidate significantly lower spin-up probability at the tone frequency starting from $\sim$~160~$\mu V_{\rm pp}$. Despite the much larger Stark shift for quantum dot qubit, this value is comparable to the $\sim$~200~$\mu V_{\rm pp}$ in the Si:P donor system, as the tone needs to overcome a $\sim$ 35 times higher noise floor before it becomes visible. Lower spin-up probabilities are also observed in the third (6.66~kHz) and fifth (4~kHz) harmonics of 20~kHz but not in the even harmonics as their effect has been suppressed by the CPMG filter function~\cite{biercuk2011dynamical}.

In summary, we have characterized and assessed the environment of a silicon quantum dot spin qubit by performing measurements of spin coherence times, Clifford-based gates randomized benchmarking, gate-induced Stark shift and noise spectroscopy. Notably, the 1-qubit control fidelity in this device is 4 times better in error rate compared to previously reported experiments even though the $T_2^*$ is 4 times shorter. We achieved this with better microwave engineering control that includes the utilization of the vector mode in our microwave source and resonance frequency feedback control. Our qubit experiences a similar noise environment as the Si:P donor qubit but we have observed significantly larger influence of $1/f$ noise in the intermediate frequency range, due to higher sensitivity of our qubits to charge noise, which results from the larger Stark shift present in quantum dot qubits. The peak at 3.6~kHz in the noise spectroscopy, found to be caused by the SIM928 voltage source, should be manageable using proper filtering techniques or alternative measurement electronics. This has emphasized the importance of noise spectroscopy measurements to probe the sources of noise that are coupled to a qubit. This experiment also highlights the capability of our quantum dot qubits as a sensitive metrological device to detect electromagnetic noise environment in a nanoelectronic circuit.

\begin{acknowledgments}
We acknowledge support from the Australian Research Council (CE11E0001017 and CE170100039), the US Army Research Office (W911NF-13-1-0024 and W911NF-17-1-0198) and the NSW Node of the Australian National Fabrication Facility. K. M. I. acknowledges support from a Grant-in-Aid for Scientific Research by MEXT, NanoQuine, FIRST, and the JSPS Core-to-Core Program.
\end{acknowledgments}

\appendix

\section{SIM928 d.c.~voltage source noise spectrum}

\begin{figure}
	\includegraphics[width=\columnwidth]{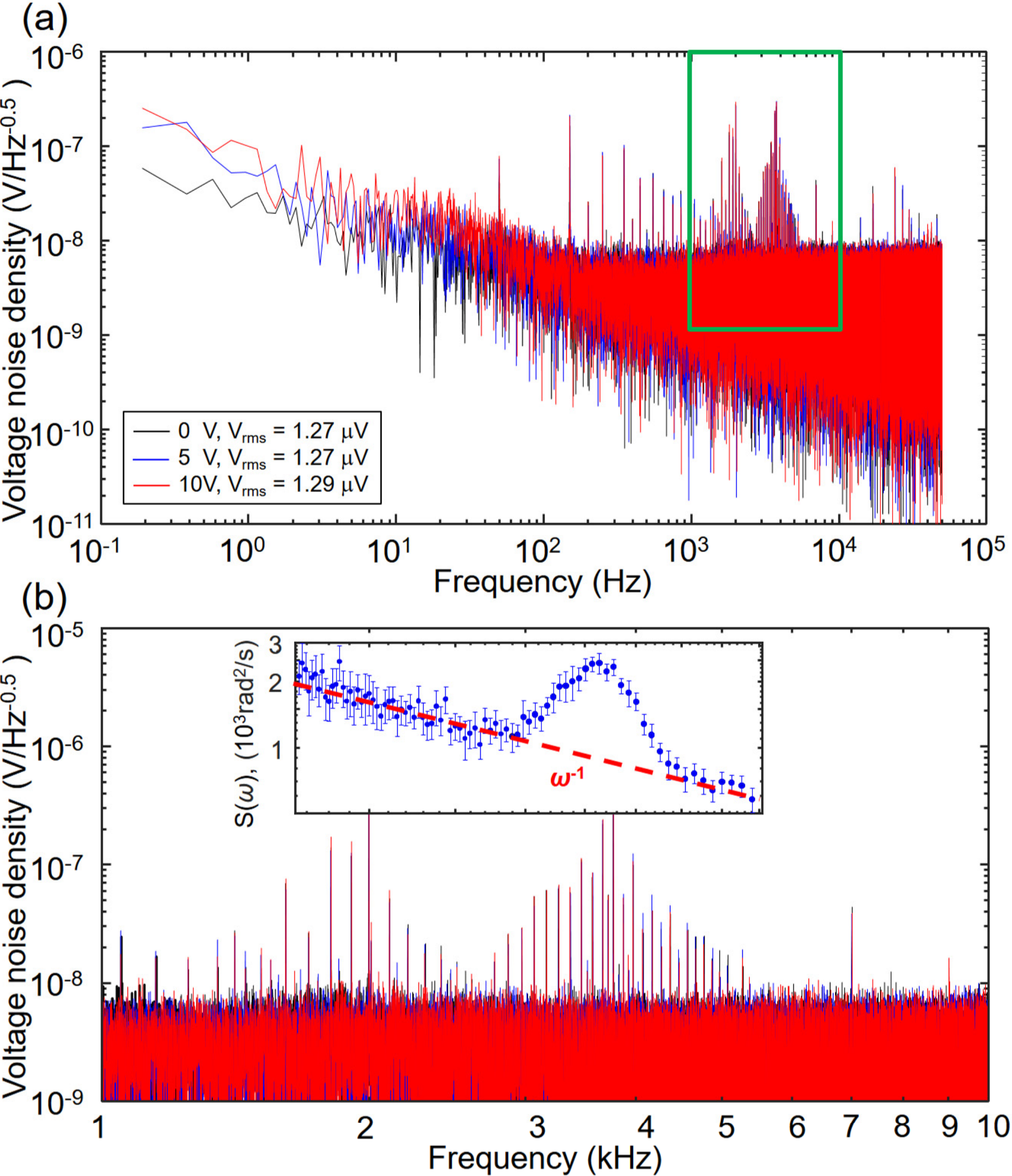}
	\caption{(a) Measured noise spectrum of the Stanford Research Systems SIM928 d.c.~voltage source used to bias the qubit device. (b) Zoom-in of the marked green region in (a) showing the noise spectrum in the 1--10~kHz range. Inset is the zoom-in noise spectroscopy of Q1, measured with more data points to exemplify the $1/f$ charge noise trend and corroborate the peak at $f \approx 3.6$~kHz. The plot is placed on the same frequency axis as the SIM928 noise spectrum, showing the matching noise peak at 3.6~kHz.} 
	\label{A2}
\end{figure}

Figure~\ref{A2}(a) is the noise spectrum of the SIM928 d.c.~voltage source used to bias the qubit device. For this measurement, the SIM928 was connected to the same type of resistive voltage divider/adder, that we use in our setups to combine d.c.~and a.c.~voltage signals. The output of the voltage adder was then fed into an SR560 voltage amplifier and recorded on a digital oscilloscope. The voltage trace is Fourier transformed to obtain the noise spectrum. The three spectra in black, blue and red are measurements taken with the SIM928 set to 0~V, 5~V and 10~V, respectively. The noise spectra are independent of the SIM928 voltage and their average fluctuations for $f$ = 0.2~Hz--50~kHz is $V_{\rm rms}$~$\approx$~1.27~$\mu V$. Figure~\ref{A2}(b) is a zoom-in of the marked green region in Fig.~\ref{A2}(a) with the inset of qubit Q1 noise spectroscopy taken at 1.5--5.5~kHz. Both plots are placed on the same frequency axis, showing the matching noise peak at 3.6~kHz.

%


\bibliography{manuscript}

\end{document}